\documentclass[aps,prb,twocolumn,groupedaddress,showpacs]{revtex4-1}
\usepackage{amsmath}
\usepackage{graphicx}
\usepackage{epstopdf}
\usepackage{bm}
\usepackage{amssymb}
\usepackage{subfigure}
\usepackage[colorlinks,linkcolor=blue,citecolor=blue]{hyperref}
\begin{document}

\title{Laser amplification in  $e^{-}$-$\mu^{-}$-$\mathrm{ion}$  plasmas}

  \author{Y. Chen$^{1}$,  R. Ou$^{1}$, H. Wang$^{1}$, S. J. Chen$^{1}$, Y. X. Zhong$^{1}$, Y. G. Chen$^{2}$, S. Tan$^{2}$, Y. X. Li$^{2}$, C. Y. Zheng$^{3,4,5}$, Z. J. Liu$^{3,4}$, L. H. Cao$^{3,4,5}$, M. M. Zhang$^{1}$, D. P. Feng$^{1}$, W. J. Zuo$^{1}$, and C. Z. Xiao$^{2,5}$}
  \email{xiaocz@hnu.edu.cn}
  \affiliation{$^1$ School of Electrical and Information Engineering, Anhui University of Science and Technology, Huainan,Anhui 232001, China}
  \affiliation{$^2$Key Laboratory for Micro-/Nano-Optoelectronic Devices of Ministry of Education, School of Physics and Electronics, Hunan University, Changsha, 410082, China}
  \affiliation{$^3$Institute of Applied Physics and Computational Mathematics, Beijing, 100094, China}
  \affiliation{$^4$\mbox{HEDPS, Center for Applied Physics and Technology, Peking University, Beijing 100871, China}}
  \affiliation{$^5$\mbox{Collaborative Innovation Center of IFSA (CICIFSA), Shanghai Jiao Tong University, Shanghai 200240,China}}
  %\date{\today}

  \begin{abstract}
   We investigate laser amplification in $e^{-}$-$\mu^{-}$-$\mathrm{ion}$ plasmas, where negative muons partially replace electrons. Theoretical results reveal a hybrid plasma wave, called $\mu$-wave that exhibits ion-acoustic behavior in long-wavelength regime and Langmuir-like behavior in short-wavelength regime. Besides, the Landau damping of $\mu$-wave is smaller than that of Langmuir wave.  Particle-in-cell (PIC) simulations confirm the theoretical results of instabilities in $e^{-}$-$\mu^{-}$-$\mathrm{ion}$  plasmas. The $\mu$-wave enables efficient laser amplification by suppressing pump-driven spontaneous instabilities through enhanced Landau damping of Langmuir waves. Compared to Raman amplification, $\mu$-wave amplification can maintain the Gaussian waveform of the seed laser, avoiding  pulse splitting. Compared to strong-coupling Brillouin amplification, $\mu$-wave amplification exhibits weaker filamentation instability. Our theoretical model can be generalized to other plasma systems containing two species of negatively charged particles, such as two-temperature electron plasmas and negative-ion plasma. These findings establish $e^{-}$-$\mu^{-}$-$\mathrm{ion}$ plasma as a promising medium for advanced laser amplification schemes.
  \end{abstract}

  \pacs{}

  \maketitle

\section{Introduction}\label{introduction}
The negative muon, a fundamental lepton with charge $-1$ and spin $1/2$, differs from the electron primarily in its larger mass ($m_\mu \approx 207m_e$). Due to their similar properties, muons have been investigated as potential electron substitutes in atomic systems\cite{muonic1,muonic2}. Recent studies have also demonstrated muon acceleration from MeV to GeV energies via plasma wakefields\cite{muonic3,muonic4}, enabling applications in muon radiography\cite{muonic5}, material probing\cite{muonic6}, and muon collider\cite{muonic7,muonic8}.

Generating high-intensity muon beams remains challenging. Current methods include: (i) cosmic-ray atmospheric interactions \cite{cosmic1,cosmic2,cosmic3}, and (ii) pion decay from proton-graphite collisions\cite{labmiu1,labmiu2,labmiu3}. More promising approaches employ laser-accelerated electrons\cite{lasermiu1,lasermiu2} or protons\cite{lasermiu4}, with recent experiments achieving 0.01 muons per electron \cite{lasermiu3}. Advances in muon yield and ionization cooling\cite{cooling} may soon enable the creation of $e^{-}$-$\mu^{-}$-$\mathrm{ion}$ plasmas. However, it must be acknowledged that present-day technology cannot generate sufficient muon yields to form a classical plasma with many particles per Debye sphere.

Recently, studies of electron-positron pair plasmas revealed enhanced Brillouin amplification \cite{e_p_2016,e_p_2017}, and our later research showed that introducing ion charge states in comparable-mass plasma leads to laser amplification enhancement\cite{chen1}. Therefore, plasma waves and instabilities in exotic plasmas need further investigations.

In this work, we report the discovery of a distinct muon-associated plasma wave in $e^{-}$-$\mu^{-}$-$\mathrm{ion}$  plasma, called $\mu$-wave. This hybrid wave exhibits ion-acoustic-like behavior in the long-wavelength limit while transitioning to Langmuir-wave-like dispersion at short wavelengths. Then we theoretically investigate laser-plasma instabilities in $e^{-}$-$\mu^{-}$-$\mathrm{ion}$  plasma, including the growth rates of both stimulated Raman scattering (SRS) and $\mu$-wave instability. The results of fully  kinetic particle-in-cell (PIC) simulations agree with theoretical predictions.

 %Pulse amplification utilizing laser-plasma instabilities represents a highly promising approach for achieving ultrahigh-intensity laser pulses.

 The conventional laser amplification schemes include Raman amplification\cite{malkin1,SRS1,SRS2,SRS3} and strong-coupling Brillouin (SC-SBS) amplification\cite{Riconda2d,SBS2,SBS4,SBS5,SBS7,SBS10,Amir}. Raman amplification has the advantage of high growth rate, but this characteristic is also prone to cause spontaneous instability of the pump laser, thereby reducing the energy conversion efficiency\cite{suppress1,suppress2,suppress3,suppress4,gas,fly}. The nonlinear frequency shift of Langmuir waves can induce the saturation and pulse splitting in Raman amplifications\cite{spliting,suppress5}. Similar to Raman amplification, SC-SBS amplification also suffers the spontaneous instabilities of pump laser\cite{suppress6,suppress7,nosie3}. In addition to this limitation, the amplification efficiency is substantially compromised by  filamentation instability \cite{SBS6,nosie1,nosie2}.

 In contrast, the efficiency of $\mu$-wave amplification (laser transfer energy through $\mu$-waves) remains largely unaffected  by spontaneous pump instabilities, owing to strong Landau damping of Langmuir waves in $e^{-}$-$\mu^{-}$-$\mathrm{ion}$  plasmas. Besides, the growth rate of filamentation instability in $e^{-}$-$\mu^{-}$-$\mathrm{ion}$  plasmas is lower than that in $e^{-}$-$\mathrm{ion}$ plasmas.  As a result, the seed laser can maintains  its Gaussian profile in $\mu$-wave amplifications. Our theoretical model for plasma waves and instabilities can be extended to other double-negative-species plasmas\cite{twoele}, providing theoretical guidance for studying laser amplification in such plasma systems.

 This paper is structured in the following ways. Firstly, in Sec.~\ref{es_waves}, we describe the electrostatic modes in $e^{-}$-$\mu^{-}$-$\mathrm{ion}$ plasma. Secondly, the growth rates of instabilities in $e^{-}$-$\mu^{-}$-$\mathrm{ion}$ plasma is discussed and perform PIC simulations to verify in Sec.~\ref{miu_grow_pic}. Thirdly, Comparison between $\mu$-wave amplification and conventional amplification schemes in Sec.~\ref{miu_com_con}. At last, the conclusion  and discussion are shown in  Sec.~\ref{conclusion}.

\section{The electrostatic waves in $e^{-}$-$\mu^{-}$-$\mathrm{ion}$ plasmas}\label{es_waves}

In traditional electron-ion plasmas, the high-frequency mode is called the Langmuir wave, while the low-frequency mode is called the ion acoustic wave. In $e^{-}$-$\mu^{-}$-$\mathrm{ion}$ plasmas,
muons bridge the mass gap between electrons and ions, so new electrostatic modes with frequencies lying between Langmuir waves and ion-acoustic waves may emerge in such plasmas. Let $\eta$ denote the electron fraction in the plasma, so electron density is $n_{e} = \eta n_{0}$ and muon density is $n_{\mu} = (1-\eta) n_{0}$, where $n_{0}$ is the density of negative charged particles.

 In order to obtain the dispersion relations of plasma waves in $e^{-}$-$\mu^{-}$-$\mathrm{ion}$ plasmas, we can write the coupling equations by using the one-dimensional two-fluid model,
  \begin{equation} \label{Bfluid_eq1}
   \begin{split}
      &\partial_{t} n_{s}+\partial_{x} (n_{s}V_{s})=0,\\
     & m_{s} n_{s}(\partial_{t} v_{s}+V_{s} \partial_{x} v_{s})=-\partial_{x} P_{s} + q_{s}n_{s}E,\\
     &\partial_{x} E=-4 \pi e (n_{\mu}+n_{e}-n_{i}),
    \end{split}
 \end{equation} where $n_{s} (s=e,\mu, i)$ are the density of each component in  $e^{-}$-$\mu^{-}$-$\mathrm{ion}$ plasma.   $V_{s} (s = e,\mu, i)$ are corresponding mean velocity of each kind of particles. Species charge states are denoted by  $Z_{s} (s = e,\mu, i) = 1$. $P_{s} (s = e,\mu, i) = \gamma_{s}T_{s}\partial_{x} n_{s}$ are  the thermal pressure for each component, where $T_{s}$ are the temperatures of particles and  $\gamma_{s}$ are the parameters to determine  whether the process is adiabatic or isothermal.

 For simplicity,  we assume that the mass of ions is infinite, they only  serve as a positive background.  Then we linearize them with $n_{s}=n_{s0}+\tilde{n}_{s}$, $V_{s}=\tilde{v}_{s}$, then we can get
 \begin{equation} \label{Bfluid_eq4}
     m_{s}\partial^{2}_{t}\tilde{n}_{s}+\partial^{2}_{x}P_{s}+q_{s}n_{s0}4\pi e(\tilde{n}_{e}+\tilde{n}_{\mu})=0.
 \end{equation}

 Use a solution with  the from $e^{i(kx-\omega t)}$  to Eq.~(\ref{Bfluid_eq4}), one can obtain,
 \begin{equation} \label{Bfluid_eq5}
\left[ \begin{array}{ccc}
a & b  \\
c &  d
\end{array}
\right ]
%\times
{\left[ \begin{array}{ccc}
 \tilde{n}_{e}  \\
 \tilde{n}_{\mu}
\end{array}
\right ]}
= 0,
 \end{equation} where $a = \omega^{2}-k^{2}C_{e}^{2}-\eta\omega_{p0}^{2}$,  $ b = -\eta\omega_{p0}^{2} $, $c = -\beta(1-\eta)\omega_{p0}^{2}$, $d = \omega^{2}-\alpha\beta k^{2}C_{e}^{2}-\beta(1-\eta)\omega_{p0}^{2}$,
 and $\omega_{p0} = \sqrt{\frac{4\pi e^{2} n_{0}}{m_{e}}}$ is the electron plasma frequency when $\eta = 1$.   $C_{e}^{2} = \frac{\gamma_{e}T_{e}}{m_{e}}$, $C_{\mu}^{2} = \frac{\gamma_{\mu}T_{\mu}}{m_{\mu}}$ and  $\beta = 1/207 $ is the mass ratio between electron and muon, $\alpha = T_{\mu}/T_{e}$ is the temperature ratio between  muon and electron. Here, we take the adiabatic assumption, $i.e.$ $\gamma_{e} = \gamma_{\mu} = 3$.

\begin{figure}[htbp]
    \begin{center}
      \includegraphics[width=0.49\textwidth,clip,angle=0]{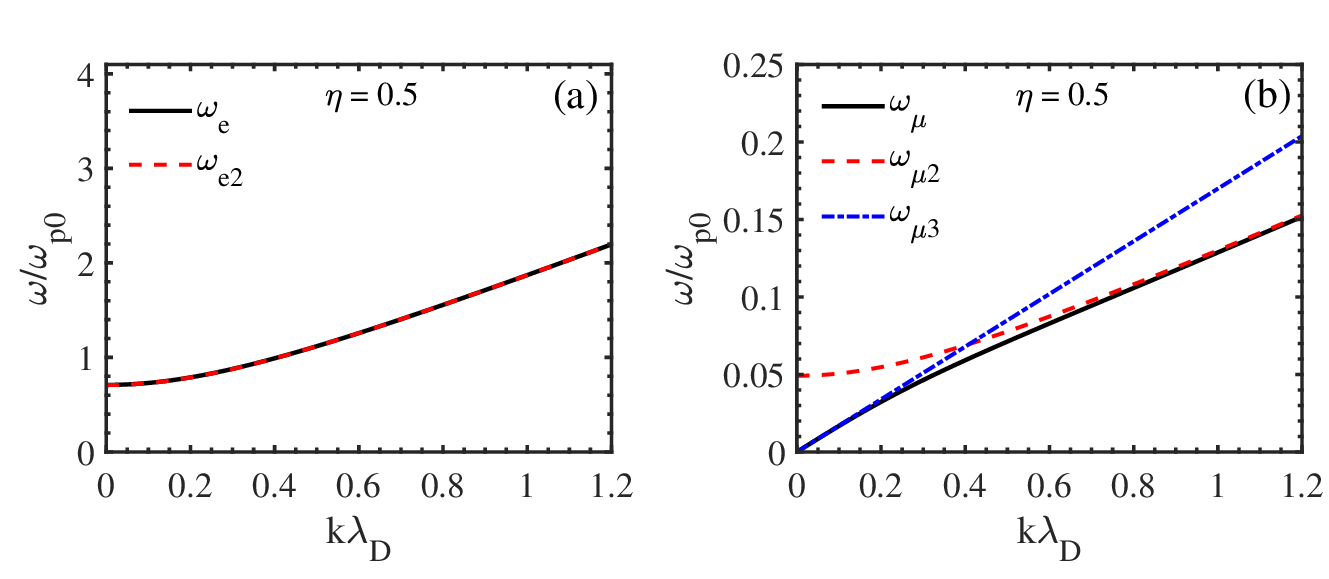}\vspace{-10pt}
      % Here is how to import EPS art [width=20mm,height=10mm][width=9cm,trim=0.5cm 21cm 0 2cm]
      \caption{\label{miu_w} (a) The frequency of electron plasma waves by two-fluid model Eq.~(\ref{eq:4a})(black straight line) and the Langmuir waves approximation solutions by Eq.~(\ref{wlw}) (red dashed line) with $\eta =0.5$. (b) The frequency of electron plasma waves by two-fluid model Eq.~(\ref{eq:4b}) (black straight line), the approximation solutions of $\mu$-wave by Eq.~(\ref{mulw}) (red dashed line) at short-wavelength regime and the approximation solutions of $\mu$-wave by Eq.~(\ref{mulw2}) (blue dotted line) at long-wavelength regime. }
    \end{center}
  \end{figure}

 For Eq.~(\ref{Bfluid_eq5}) to possess a non-zero solution, it is necessary that the determinant of its coefficient matrix equals to $0$. Solving Eq.~(\ref{Bfluid_eq5}), one can obtain the resonant frequency of plasma waves,
\begin{subequations}
\label{eq:4a}
\begin{align}
  \label{eq:4a} % Langmuir wave branch
  \omega_{e}^{2} &= \frac{\omega_{ek}^{2}}{2} + \frac{\sqrt{\omega_{ek}^{4}-4\beta k^{2}C_{e}^{2}[\alpha k^{2}C_{e}^{2}+(1- \eta+\eta\alpha)\omega_{pe0}^{2}]}}{2}, \\
  \label{eq:4b} % Mu-wave branch
  \omega_{\mu}^{2} &= \frac{\omega_{ek}^{2}}{2} - \frac{\sqrt{\omega_{ek}^{4}-4\beta k^{2}C_{e}^{2}[\alpha k^{2}C_{e}^{2}+(1- \eta+\eta\alpha)\omega_{pe0}^{2}]}}{2}.
\end{align}
\end{subequations} where $\omega_{ek}^{2} = (\beta- \beta \eta +\eta)\omega_{p0}^{2} + (1+\alpha\beta)k^{2}C_{e}^{2}$, the $\omega_{e}$ corresponds to the Langmuir wave, and $\omega_{\mu}$ corresponds to the $\mu$-wave. The fluid model used here is valid in the long-wavelength limit, i.e., when the wave number satisfies $k^{2}\lambda_{Ds}^{2} \ll 1$, where $\lambda_{De}$ and $\lambda_{D\mu}$ are the electron and muon Debye lengths, respectively.  The  $\mu$-wave branch to be a well-defined normal mode, the plasma must contain sufficient particles of both types to support the oscillation. These criteria imply that the  neither the electron fraction nor the muon fraction can be too small for a given wave number. Besides,  In the long-wavelength limit ($k \rightarrow 0$), the $\mu$-wave branch described by Eq.~(\ref{eq:4b}) exhibits ion-acoustic behavior.

For intermediate cases ($0 < \eta < 1$), The dispersion relation of Langmuir wave branch can be approximated by
\begin{equation} \label{wlw}
     \omega_{e2}^{2} = \eta\omega_{p0}^{2} + 3k^{2}v_{e}^{2}.
 \end{equation}

 The black straight line in Fig.~\ref{miu_w}(a) is the exact dispersion relations of Langmuir waves for  $\eta = 0.5$, and it agrees with the approximate solution of Langmuir waves (red dashed lines) by Eq.~(\ref{wlw}) when we treat muons as immobile particles, which means muon density perturbations barely affect Langmuir wave dispersion.

 The situation becomes different when discussing the dispersion relation of $\mu$-waves. At the short-wavelength regime (large $k \lambda_{D}$), where $ \lambda_{D} =  v_{e}/\omega_{p0}$ are the Debye length, the muon density perturbations are decoupled from electron density perturbations. The behavior of muons is similar to electrons, so the approximate solution of $\mu$-waves can be obtained,
\begin{equation} \label{mulw}
      \omega_{\mu 2}^{2} = (1-\eta)\beta\omega_{p0}^{2} + 3k^{2}v_{\mu}^{2}.
 \end{equation}

In Fig.~\ref{miu_w}(b), the result of approximate solutions of $\mu$-waves (red dashed line) agree with Eq.~(\ref{eq:4b}) (black straight line ) at large $k \lambda_{D}$. However the approximate solution Eq.~(\ref{mulw}) breaks down in the long-wavelength regime, The dispersion relations of $\mu$-wave have the characteristics of ion acoustic waves at long-wavelength limit. Because at long wavelengths, just like in ion acoustic waves where electrons create pressure and ions carry inertia; in $e^{-}$-$\mu^{-}$-$\mathrm{ion}$  plasmas,  electrons provide thermal pressure while the heavier muons move slowly due to their mass. In order to obtain the approximate solution of $\mu$-waves in long-wavelength regime, one can take the approximation $k\rightarrow 0$,  for the second term of Eq.~(\ref{eq:4b}), one can Drop the $k^{4}$ terms and take the Taylor expansion at $k^{2} = 0$,  Eq.~(\ref{eq:4b}) becomes,
\begin{equation} \label{mulw3}
      \omega_{\mu3 }^{2} = \frac{\omega_{ek}^{2}}{2} - \frac{1}{2}\left[\omega_{ek}^{2}-\frac{4\beta(1-\eta+\eta \alpha)k^{2}C_{e}^{2}\omega_{p0}^{2}}{\omega_{ek}^{2}}\right].
 \end{equation}

Finally, the approximate solution of $\mu$-waves at low $k\lambda_{D}$ regime is,
\begin{equation} \label{mulw2}
\begin{split}
      &\omega_{\mu 3} = k v_{\phi}(\eta), \\
     & v_{\phi}(\eta) = C_{\mu}\sqrt{\frac{(1-\eta+\eta\alpha)}{\alpha(\beta- \beta \eta +\eta)}},
\end{split}
 \end{equation} here, the temperature ratio $\alpha = 1$, so the phase velocity of $\mu$-wave becomes,
\begin{equation} \label{mulw4}
  v_{\phi}(\eta) = C_{\mu}\sqrt{\frac{1}{\beta- \beta \eta +\eta}}.
\end{equation}

As shown in Fig.~\ref{miu_w}(b), the approximate solution of $\mu$-waves (blue doted line) agrees well with the solution by  Eq.~(\ref{eq:4b})  at low $k\lambda_{D}$ regime. The properties of $\mu$-waves closely resemble those of electron acoustic waves (EAWs) in two-electron-temperature plasmas\cite{twoele,twoele2,twoele3}. Our theoretical model can be generalized to such plasmas by adjusting the parameters $\beta$ (effective mass ratio) and $\alpha$ (temperature ratio).

 \begin{figure}[htbp]
    \begin{center}
      \includegraphics[width=0.49\textwidth,clip,angle=0]{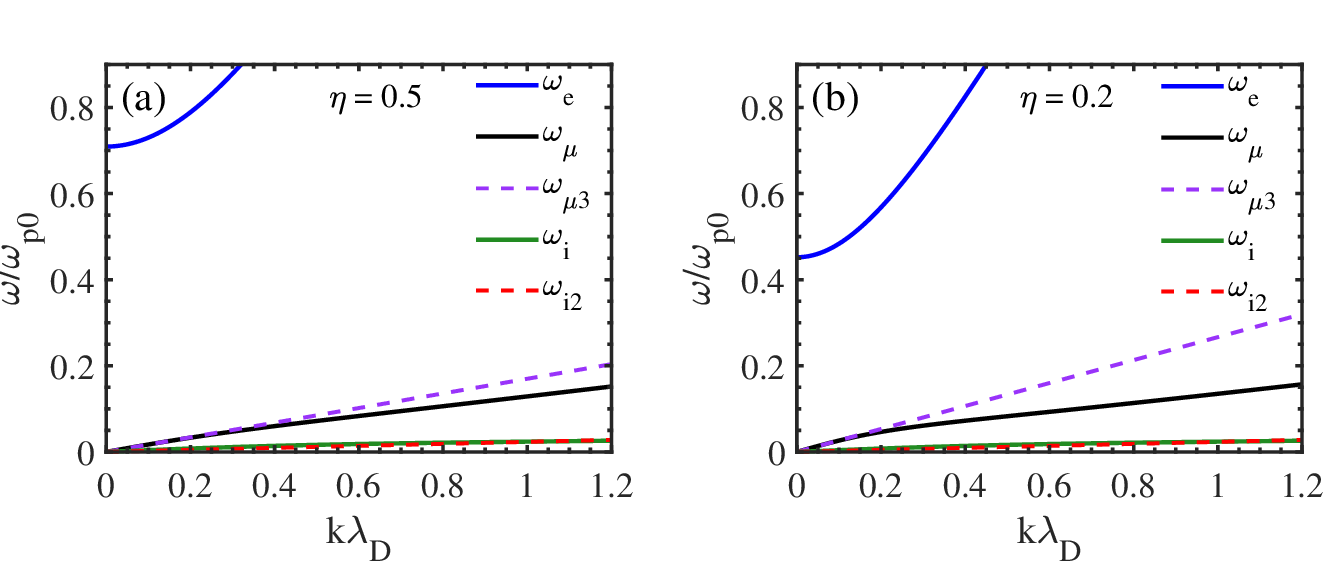}\vspace{-10pt}
      % Here is how to import EPS art [width=20mm,height=10mm][width=9cm,trim=0.5cm 21cm 0 2cm]
      \caption{\label{ion_w} (a)The dispersion relations of three branches of wave modes when $\eta = 0.5$, (b)The dispersion relations of three branches of wave modes when $\eta = 0.2$. Langmuir wave (blue straight lines), $\mu$-wave (black straight lines) and ion acoustic wave(green straight lines ), they are obtained by numerically solving  Eq.~(\ref{Bfluid_eq15}). The purple dashed lines are the approximate solution of $\mu$-wave by Eq.~(\ref{mulw2}), red dashed lines are the approximate solution of ion acoustic wave by  Eq.~(\ref{threecom2}). }
    \end{center}
  \end{figure}

 For limiting cases, when $\eta = 1$ (pure electron plasma), $\omega_{ek}^{2} = \omega_{p0}^{2}+(1+\alpha\beta)k^{2}C_{e}^{2}$, Substituting $\omega_{ek}^{2}$ to Eq.~(\ref{eq:4a}), one can obtain the dispersion relation of Langmuir wave, $\omega_{e}^{2} = \omega_{p0}^{2} + 3k^{2}v_{e}^{2}$, the $\mu$-wave solution $\omega_\mu$ becoming physically irrelevant. When $\eta = 0$ (pure muon plasma), we use the similar treatment, then Eq.~(\ref{eq:4b}) becomes $\omega_{\mu}^{2} = \beta\omega_{p0}^{2} + 3k^{2}v_{\mu}^{2}$, with the electron solution $\omega_{e}$ becoming meaningless, where $v_{e}$ and $v_{\mu}$ are the thermal velocity of electron and muon, respectively.

The model we adopted in the derivation above neglects ion motion. This approximation holds when the ion mass is significantly greater than that of the muon. When ions are protons, the muon mass is only about 11\% of the proton mass, one might hypothesize that the wave modes in a muon-proton plasma would couple under these conditions, to definitively show that the proton motion does not significantly affect the dispersion relation of the muon wave, the equations of motion for both species must be solved simultaneously to obtain an accurate dispersion relation.

If we consider the motion of protons, Eq.~(\ref{Bfluid_eq4}) becomes,
\begin{equation} \label{Bfluid_eq10}
     m_{s}\partial^{2}_{t}\tilde{n}_{s}+\partial^{2}_{x}P_{s}+q_{s}n_{s0}4\pi e(\tilde{n}_{e}+\tilde{n}_{\mu}-\tilde{n}_{p})=0.
 \end{equation}

 Then we assume a solution with  the from $e^{i(kx-\omega t)}$  to Eq.~(\ref{Bfluid_eq10}), one can obtain,
\begin{equation} \label{Bfluid_eq15}
\left[ \begin{array}{ccc}
a & b &  A_{e} \\
c & d &  A_{\mu} \\
A_{p} & A_{p} &  \omega^{2}-B_{p}-A_{p}
\end{array}
\right ]
%\times
{\left[ \begin{array}{ccc}
 \tilde{n}_{e}  \\
 \tilde{n}_{\mu}\\
 \tilde{n}_{p}
\end{array}
\right ]}
= 0,
\end{equation} where $A_{e} = \eta\omega_{p0}^{2}$, $A_{\mu} = \beta(1-\eta)\omega_{p0}^{2}$, $A_{p} = \beta_{p}\omega_{p0}^{2}$ and $B_{p} = 3k^{2}\alpha_{p}\beta_{p} C_{e}^{2}$,  where $\alpha_{p}=T_{i}/T_{e}$ is the temperature ratio of ions and electrons, $\beta_{p}=m_{e}/m_{p}$ is the mass ratio of ions and electrons.

For Eq.~(\ref{Bfluid_eq15}) to possess a non-zero solution, it is necessary that the determinant of its coefficient matrix equals to $0$. The analytical solutions are exceedingly complex; therefore, we employed numerical methods to solve the equations and obtained the dispersion relations for the three types of waves.

By numerically solving Eq.~(\ref{Bfluid_eq15}) with $\eta = 0.5$ and $\eta = 0.2$, we obtained the numerical solution of plasma waves in $e^{-}$-$\mu^{-}$-ion plasmas.  As shown in Fig.~\ref{ion_w}, blue straight lines, black straight lines and green straight lines are the dispersion relations of Langmuir wave, $\mu$-wave and ion acoustic wave, respectively. Purple dashed lines are the approximate solution of $\mu$-wave by Eq.~(\ref{mulw2}), they agree with numerical solution of $\mu$ wave (black straight lines) at long wavelength regime, they also agree with the dispersion relation when we ignore the ion motions shown in Fig.~\ref{miu_w}(b). Therefore, the stationary ion approximation adopted in Eq.~(\ref{eq:4a}) and Eq.~(\ref{eq:4b}) is reasonable.

 Now, we turn to obtain the dispersion relation of ion acoustic waves in $e^{-}$-$\mu^{-}$-$\mathrm{ion}$  plasmas, from Eq.~(\ref{Bfluid_eq15}) by setting the determinant of matrix to zero and then making the standard assumption for ion acoustic waves: the frequencies of interest are much lower than the electron and muon plasma frequencies, we can write the dispersion relation equation,
\begin{equation} \label{threecom}
  1-\frac{\eta\omega_{p0}^{2}}{\omega^{2}-3k^{2}v_{e}^{2}}-\frac{(1-\eta)\omega_{p0}^{2}}{\omega^{2}-3k^{2}\alpha\beta v_{e}^{2}}-\frac{\beta_{p}\omega_{p0}^{2}}{\omega^{2}-3k^{2} \alpha_{p}\beta_{p} v_{e}^{2}} = 0,
\end{equation}

If we assume the temperature of ion is much lower than $T_{e}$ and $T_{\mu}$, $i.e.$ $T_{i}= 0.1T_{e}$.  The approximate solution of  ion acoustic wave is given by,
\begin{equation} \label{threecom2}
 \omega_{i2} = k \sqrt{\frac{(n_{e}+n_{\mu})T_{e}T_{\mu}}{m_{p}(n_{e}T_{\mu}+n_{\mu}T_{e})}}.
\end{equation}

As shown in Fig.~\ref{ion_w}, red dashed lines are obtained by Eq.~(\ref{threecom2}), they agree well with the numerical solution of ion acoustic wave(green straight lines).

The Eq.~(\ref{threecom2}) agrees with the dispersion relation of ion acoustic wave in electron-ion plasmas when $n_{\mu} = 0$, and it can be generalized to ion acoustic waves in two-temperature electron plasmas\cite{twoele}.

\section{Growth Rates of Laser-Plasma Instabilities and Verification via PIC Simulation}\label{miu_grow_pic}

 Laser plasma interactions in the $e^{-}$-$\mu^{-}$-$\mathrm{ion}$ plasmas need to be discussed , because the $\mu$-wave has potential applications in laser amplifications. Following the treatments in  early works\cite{e_p_2016,chen1}, if we consider mass of ion is infinite, the coupling equations for the instabilities can be written,
\begin{equation} \label{gw5}
\begin{split}
&[\partial_{t}^{2}-c^{2}\nabla^{2}+(\eta+\beta-\beta \eta)\omega_{pe}^{2}]A_{seed}=-\frac{4\pi e^{2}}{m_{e}}
\tilde{n} A_{pump},\\
& (\partial_{t}^{2}-C_{e}^{2}\nabla^{2})\tilde{n}_{e}+\eta\omega_{pe}^{2}(\tilde{n}_{e}+\tilde{n}_{\mu})=\frac{tn_{e0}e^{2}}{m_{e}^{2}c^{2}}\nabla^{2}(A_{pump}\cdot A_{seed}),\\
   &(\partial_{t}^{2}- C_{\mu}^{2}\nabla^{2})\tilde{n}_{\mu}-(1-\eta)\beta\omega_{pe}^{2}(\tilde{n}_{e}+\tilde{n}_{\mu})=\frac{\beta^{2}(1-\eta)n_{e0}e^{2}}{m_{e}^{2}c^{2}} \\
   &\nabla^{2}(A_{pump}\cdot A_{seed}),
\end{split}
  \end{equation} where $A_{pump}$ and $A_{seed}$ represent the vector potentials of the pump laser and seed laser, respectively. $\tilde{n} = \tilde{n}_{e}+\beta \tilde{n}_{\mu}$ is the total density fluctuation.

We Fourier-analyze these equations,
\begin{equation} \label{fA17}
\left[ \begin{array}{ccc}
a & b  \\
c &  d
\end{array}
\right ]
%\times
{\left[ \begin{array}{ccc}
 \tilde{n}_{e}  \\
 \tilde{n}_{\mu}
\end{array}
\right ]}
=-{\left[ \begin{array}{ccc}
 \eta  \\
 \beta^{2}(1-\eta)
\end{array}
\right ]}F,
 \end{equation} where $F=
   (k^{2}v_{osc}^{2}\omega_{p0}^{2} )(\tilde{n}_{e}+\beta \tilde{n}_{\mu})/4D(k-k_{pump},\omega-\omega_{pump})$,where $D(k,\omega) = \omega^{2} - k^2c_{0}^2-[\eta+\beta(1-\eta)]\omega_{p0}^{2} $, where $c_0$ is the speed of light in vacuum.  We can also write  Eq.~(\ref{fA17})as
\begin{equation} \label{Agw24}
 \begin{split}
  &\left[(\omega-\omega_{pump})^{2}-(k-k_{pump})^{2}c_{0}^{2}-(\eta+\beta-\beta \eta)\omega_{p0}^{2}\right]\\
  &(ad-bc)=\frac{k^{2}v_{osc}^{2}\omega_{p0}^{2}}{4}(\eta d-\beta \eta c + a \beta^{3}-\\
  &\eta a\beta^{3}-b\beta^{2}+ \eta b\beta^{2}).
 \end{split}
 \end{equation}

   Assuming $\omega = \omega_{e,\mu} + \delta$ and substituting it into Eq.~(\ref{Agw24}) with $\delta \ll \omega_{e,\mu}$.  We only keep the terms about $\delta^2$,
 \begin{equation} \label{Agw26}
 \delta^{2}(\omega_{e,\mu}-\omega_{pump})\omega_{e,\mu} \xi_{2}
=\frac{\omega_{pe}^{2}k^{2}v_{osc}^{2}}{16}\xi_{1},
 \end{equation} where $\xi_{1} = [\eta+\beta^{3}(1-\eta)]\omega_{e,\mu}^{2}-[\eta\alpha\beta+(1-\eta)\beta^{3}]k^{2}C_{e}^{2}-[\beta \eta(1-\eta)(1+\beta)^{2}]\omega_{p0}^{2}$, $\xi_{2} = 2\omega_{e,\mu}^{2}-(1+\alpha\beta)k^{2}C_{e}^{2}-(\eta+\beta-\beta \eta)\omega_{p0}^{2}$.

    Finally, one can  obtain the growth rates of instabilities in $e^{-}$-$\mu^{-}$-$\mathrm{ion}$ plasmas,
\begin{equation} \label{fA19}
  \Gamma_{e,\mu}=\frac{kv_{osc}}{4}\sqrt{\frac{\xi_1}{\xi_2}}\left[ \frac{\omega_{p0}^{2}}{(\omega_{pump}-\omega_{e,\mu})\omega_{e,\mu}}\right]^{1/2}
  \end{equation}  $\Gamma_{e}$ is the growth rate of Raman instability, $\Gamma_{\mu}$ is the growth rate of Muon instability. If $\eta = 1$, Eq.~(\ref{fA19}) recover the growth rate of Raman instability in electron-ion plasmas.

  In this section, we demonstrate that when protons are selected as the ionic species, the impact of ion motion on the growth rate is negligible. When we consider ion motion,the couple equations becomes
\begin{equation} \label{fA20}
\left[ \begin{array}{ccc}
a & b &  A_{e} \\
c & d &  A_{\mu} \\
A_{p} & A_{p} &  s
\end{array}
\right ]
%\times
{\left[ \begin{array}{ccc}
 \tilde{n}_{e}  \\
 \tilde{n}_{\mu}\\
 \tilde{n}_{p}
\end{array}
\right ]}
= -{\left[ \begin{array}{ccc}
 \eta  \\
 \beta^{2}(1-\eta)\\
 \beta_{p}^{2}
\end{array}
\right ]}F^{(1)},
\end{equation}where $ s = \omega^{2}-B_{p}-A_{p}$, $F^{(1)}=\frac{n_{0}e^{2}k^{2}}{m_{e}^{2}c^{2}}A_{pump}\cdot A_{seed}$ , By using Schur complement elimination, we can incorporate the effects of ion motion while still maintaining the second-order nature of the equations. Finally, we get the growth rates of instabilities with considering the ion motion,
\begin{equation}\label{fA21}
  \tilde\Gamma_{e,\mu}=\frac{k v_{\rm osc}}{4}\,
  \sqrt{\frac{\tilde\xi_1}{\tilde\xi_2}}\;
  \left[\frac{\omega_{p0}^2}{(\omega_{\rm pump}-\omega_{e, \mu})\,\omega_{e, \mu}}\right]^{1/2},
\end{equation}where $\tilde\xi_1,\tilde\xi_2$ are the modified algebraic combinations. Expanding to first order in $\frac{A_p}{s}$ we obtain
\begin{equation}\label{eq:xi_expand}
  \tilde\xi_j(\omega,k)=\xi_j+\Delta\xi_j,\qquad
\end{equation}
with $\xi_j$ the immobile-ion expressions given in Eq.~(\ref{fA19}), where the correction terms are
\begin{equation}\label{correct_term}
 \begin{split}
 & \Delta\xi_1
  =\frac{\beta_p\,\omega_{p0}^2}{s}\,
    \beta \eta (1-\eta) (1-\beta)^{2}\\
 &  \Delta\xi_2 = \frac{\beta_p\,\omega_{p0}^2}{s}  2\beta \eta (1-\eta).
 \end{split}
\end{equation}

\begin{figure}[htbp]
    \begin{center}
      \includegraphics[width=0.49\textwidth,clip,angle=0]{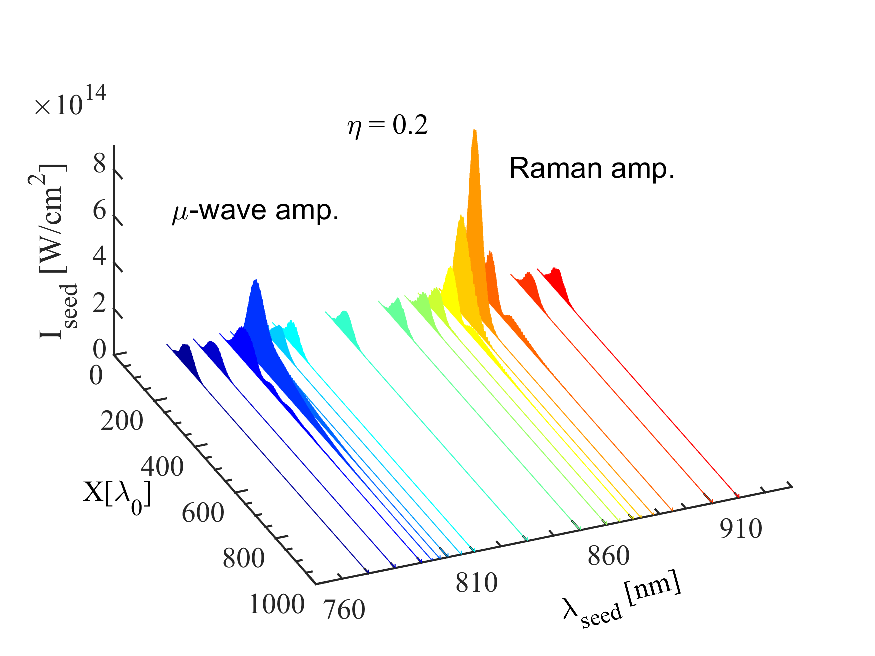}\vspace{-10pt}
      % Here is how to import EPS art [width=20mm,height=10mm][width=9cm,trim=0.5cm 21cm 0 2cm]
      \caption{\label{miu_amp} One dimensional PIC results, the seed lasers after amplifications with electron fraction $\eta = 0.2$. Cases where the seed laser wavelength is approximately $804$ $\rm{nm}$ belong to $\mu$-wave amplification, while cases with a seed laser wavelength around $880$ $\rm{nm}$ belong to Raman amplification.     }
    \end{center}
  \end{figure}

We can observe that the order of magnitude of  $\xi_j$ is $\omega_{p0}^{2}$,  while the order of magnitude of $\Delta\xi_{j}$ is $\beta \beta_{p} \omega_{p0}^{2}$, so $\Delta\xi_j \ll \xi_j$. Expanding the square root to first order in $\Delta\xi_j$ gives
\begin{equation}\label{expand22}
 \begin{split}
  &\sqrt{\frac{\tilde\xi_1}{\tilde\xi_2}}
  =\sqrt{\frac{\xi_1+\Delta\xi_1}{\xi_2+\Delta\xi_2}}  \\
  &=\sqrt{\frac{\xi_1}{\xi_2}}\left[1+\frac{1}{2}\left(\frac{\Delta\xi_1}{\xi_1}-\frac{\Delta\xi_2}{\xi_2}\right)+O(\beta_p^2)\right],
  \end{split}
\end{equation} the relative correction is
\begin{equation}\label{eq:Gamma_relative}
  \frac{\tilde\Gamma-\Gamma}{\Gamma}
  \approx \frac{1}{2}\left(\frac{\Delta\xi_1}{\xi_1}-\frac{\Delta\xi_2}{\xi_2}\right),
\end{equation} for protons, $\beta_p\approx 1/1836$, hence the relative correction is negligible ($ < 0.1\%$). That means Eq.~(\ref{fA19}) is valid when we use protons as the ions.

\begin{figure}[htbp]
    \begin{center}
      \includegraphics[width=0.49\textwidth,clip,angle=0]{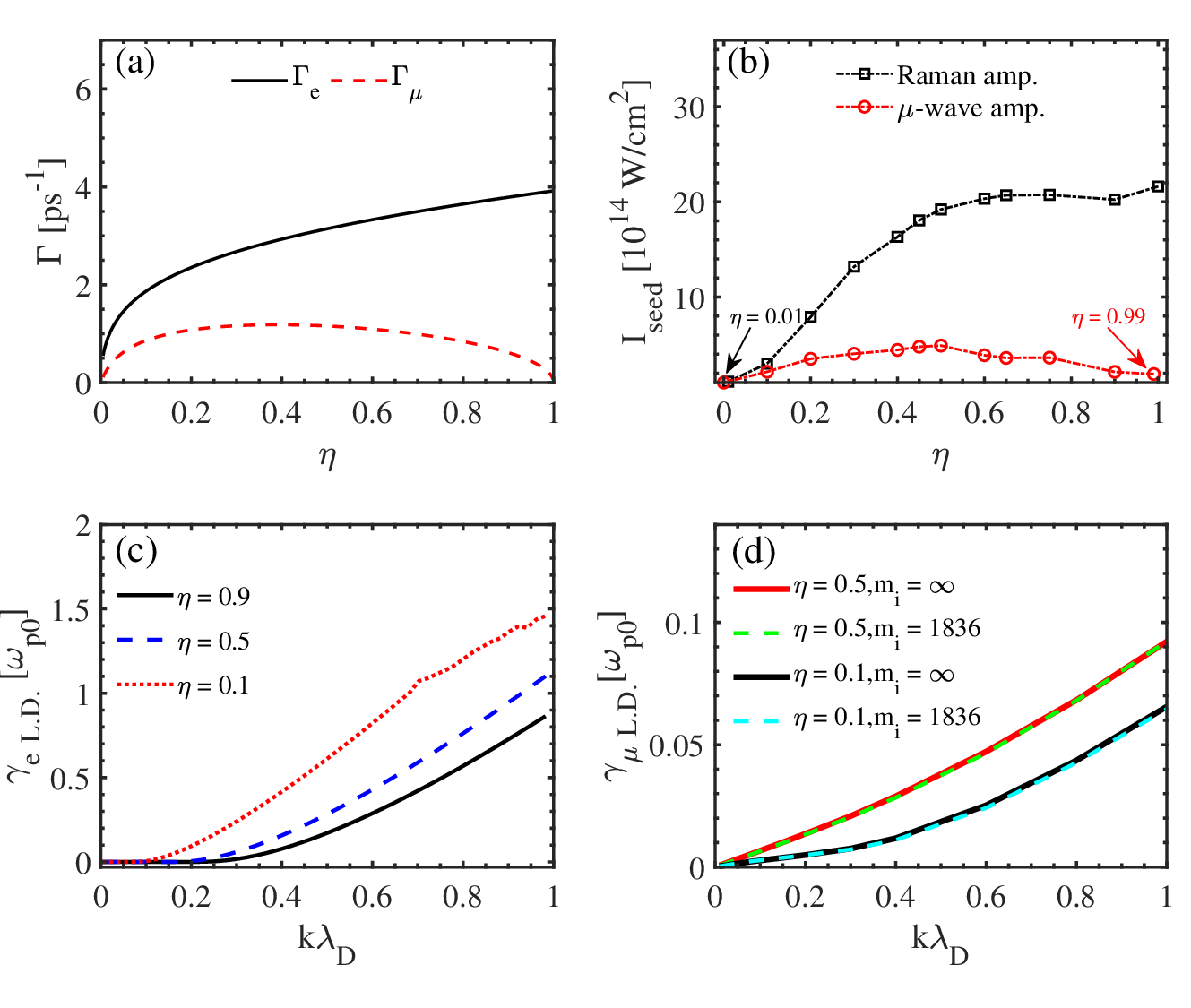}\vspace{-10pt}
      % Here is how to import EPS art [width=20mm,height=10mm][width=9cm,trim=0.5cm 21cm 0 2cm]
      \caption{\label{grow_pic} (a) The growth rate of laser plasma instabilities in electron-muon plasmas by Eq.~(\ref{fA19}), the black line is the growth rate of Raman scattering with different $\eta$, the red dashed line is the growth rate of $\mu$-wave scattering with different $\eta$  (b) PIC simulation results, the seed laser intensities for Raman amplifications (black squares) and $\mu$-wave amplifications (red circles) with different $\eta$. Both simulation and theoretical results demonstrate consistent trends. (c) The Landau damping of Langmuir waves for different $\eta$. (d) The Landau damping of $\mu$-waves for different $\eta$. }
    \end{center}
  \end{figure}

  The muon has a lifetime of approximately 2.2 microseconds, decaying into an electron, a neutrino, and an antineutrino. The timescale of the $\mu$-wave amplification interaction is on the order of picoseconds. Under ideal conditions, the muon lifetime amply satisfies the temporal requirements for muon-wave amplification. Nevertheless, achieving this still demands extremely stringent experimental conditions.

 To verify our theoretical predictions, we perform PIC simulations of pulse amplification dynamics using the \textsc{epoch} code. This computational approach has previously proven effective in studying electron-positron pair instabilities. The simulation setup included a 0.8 mm plasma region, with the plasma density $n_{0} = n_{e}+n_{\mu}= 6.9\times 10^{19} $ $\rm{cm^{-3}}$, where $n_{e} = t n_{e0}$. The ion in plasma is proton and the density of proton is equal to $n_{0}$. The temperatures of electrons, muons and protons are all $70$ $\rm{eV}$. There are $80$ cells$/\lambda_{0}$ and $100$ particles per cell in PIC simulations, where $\lambda_{0} = 800 \rm{nm}$ is the wavelength of pump laser. We obtain $k\lambda_{D} = 0.1171$ for plasma waves in these PIC simulations, then we can clearly know that $\mu$-waves exhibit characteristics of ion acoustic waves under these conditions as shown in Fig.~\ref{miu_w} (b).

A flat-top pump laser is injected from the left boundary of the simulation box, while a Gaussian seed laser pulse with a full width at half maximum (FWHM) of 113 fs enters from the right boundary. The intensities of pump laser and seed laser are both $ 10^{14}$ $\rm{W/cm^2}$. These laser beams interact with the plasma waves, exciting a three-wave instability. By changing the wavelength of seed laser,  two distinct laser amplification mechanisms exit: (i) SRS-driven Raman amplification, utilizing Langmuir waves for energy transferring, and (ii) $\mu$-wave amplification, where energy is transferred via $\mu$-waves. As shown in Fig.~\ref{miu_amp}, we scan the amplifications of different seed laser wavelengths under the condition of $\eta = 0.2$, The results show that the resonant wavelength of the seed light is 804 nm in the $\mu$-wave amplification regime, while it is 888 nm in the Raman amplification regime. The resonant  wavelengths of seed laser agree with theoretical predications by $800 \rm{nm}/(1-\omega_{e,\mu}/\omega_{pump})$ for phase-matching conditions.

 Based on the theoretical predictions shown in   Fig.~\ref{grow_pic} (a), The growth rate of stimulated Raman scattering (SRS) monotonically decreases with the electron fraction $i.e.$ $\eta$ decreases from $1$ to $0.01$, whereas the growth rate of $\mu$-wave scattering exhibits a non-monotonic trend, peaking at $\eta = 0.45$. The PIC simulation results share same trends with theoretical results, as shown in Fig.~\ref{grow_pic} (b).

 Our PIC simulations in Fig.~\ref{grow_pic}(b) show that for $\eta <0.5$ the Raman light intensity decreases faster with decreasing $\eta$ than predicted by fluid theory Fig.~\ref{grow_pic} (a), This discrepancy arises because the theory neglects Landau damping. To analyze this, we derive the kinetic dispersion relation for a multi-species plasma from the dielectric function\cite{xiehs}:
\begin{equation} \label{kDL_1}
  \epsilon(\omega,k) = 1+\chi_{e} + \chi_{\mu} + \chi_{i},
 \end{equation} where $\chi_{s} = \frac{1}{(k\lambda_{Ds})^{2}} (1+\zeta_{s}Z(\zeta_{s})), (s = e,\mu, i)$ are the electron, muon, and ion susceptibilities, respectively. where $\zeta_{s}=\omega/\sqrt{2}kv_{s}$, $\lambda_{Ds}$ are the Debye length of electron muon and ion, respectively.  $Z(\zeta)$ is the plasma dispersion function \cite{xiehs}.

 For immobile ions ($\chi_{i} \approx 0$ ), Eq.~(\ref{kDL_1}) reduces to,
 \begin{equation} \label{kDL_2}
  \epsilon(\omega,k) = 1+\chi_{e} + \chi_{\mu}.
 \end{equation} Solving for $\epsilon(\omega,k) = 0$ with $\omega = \omega_{r} - i\gamma_{e\mu L.D.} $, we find that $\gamma_{e\mu L.D.}$ (the Landau damping) increases as $\eta$ decreases (Fig.~\ref{grow_pic}(c)) explaining the faster decrease  in simulations. Further, solving Eq.~(\ref{kDL_1}) and Eq.~(\ref{kDL_2}) for proton and infinite-mass ion  reveal that the Landau damping of $\mu$-waves decreases with increasing $\eta$ (Fig.~\ref{grow_pic}(d)). This mechanism is analogous to operating a traditional Raman amplifier in a high-temperature regime, where increased Landau damping also suppresses SRS. In our case, we achieve this stabilizing effect not by globally heating the plasma, but by introducing a particle species that modifies the kinetic properties of the plasma.

 Additionally, $\mu$-waves damping grows with $k\lambda_{D}$  suggesting that operating in the low-$k\lambda_{D}$ regime optimizes amplification efficiency.

\begin{figure}[htbp]
    \begin{center}
      \includegraphics[width=0.49\textwidth,clip,angle=0]{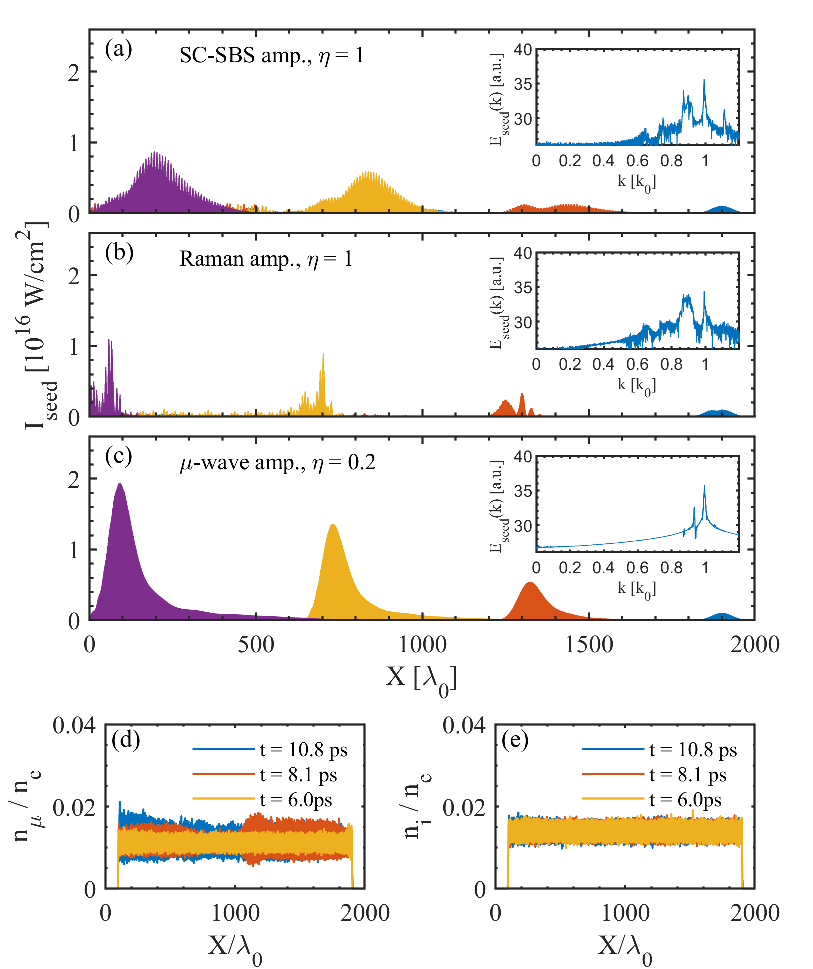}\vspace{-10pt}
      % Here is how to import EPS art [width=20mm,height=10mm][width=9cm,trim=0.5cm 21cm 0 2cm]
      \caption{\label{scamp}
      (a) The SC-SBS amplification in electron-ion plasma by PIC simulation.(b)The Raman amplification in electron-ion plasma by PIC simulation. (c)The $\mu$-wave amplification in muon-containing plasma with $\eta = 0.2$. The inset diagram displays the wavevector spectrum corresponding to the seed laser.  (d) The density of muons in $\mu$-wave amplification with $\eta = 0.2$. (d) The density of protons in $\mu$-wave amplification with $\eta = 0.2$. }
    \end{center}
  \end{figure}

\section{Comparison Between $\mu$-Wave Amplification and Conventional Amplification Schemes}\label{miu_com_con}

The $\mu$-wave  amplification  demonstrates higher energy conversion efficiency than conventional Raman techniques, owing to its lower-frequency $\mu$-wave enabling more efficient pump-to-seed photon energy transfer. Also, $\mu$-wave amplification achieves larger growth rates compared to SC-SBS amplification, as the muon mass ($m_{\mu} \simeq 207 m_e$) is significantly smaller than typical ion masses ($m_i \geq 1836 m_e$), However, as clearly shown in Fig.~\ref{grow_pic} (a), the growth rate of the $\mu$-wave instability is lower than that of SRS.  Moreover, in conventional schemes, the amplification efficiencies of both Raman and SC-SBS processes are limited by spontaneous instabilities induced by the pump laser, which parasitically deplete pump energy. To mitigate this issue, several approaches have been proposed, including: chirped pump lasers\cite{suppress6,suppress7}, inhomogeneous plasma profiles\cite{SBS6,suppress5,SBS10,Amir}, gas medium\cite{suppress2,suppress3,suppress4,gas} and flying focus techniques\cite{fly,fly2}. Now, we aim to demonstrate that $\mu$-wave amplification can also suppress pump's spontaneous instabilities.

\begin{figure}[htbp]
    \centering
    \begin{minipage}[t]{\linewidth}
        \centering
        \includegraphics[width=0.95\linewidth]{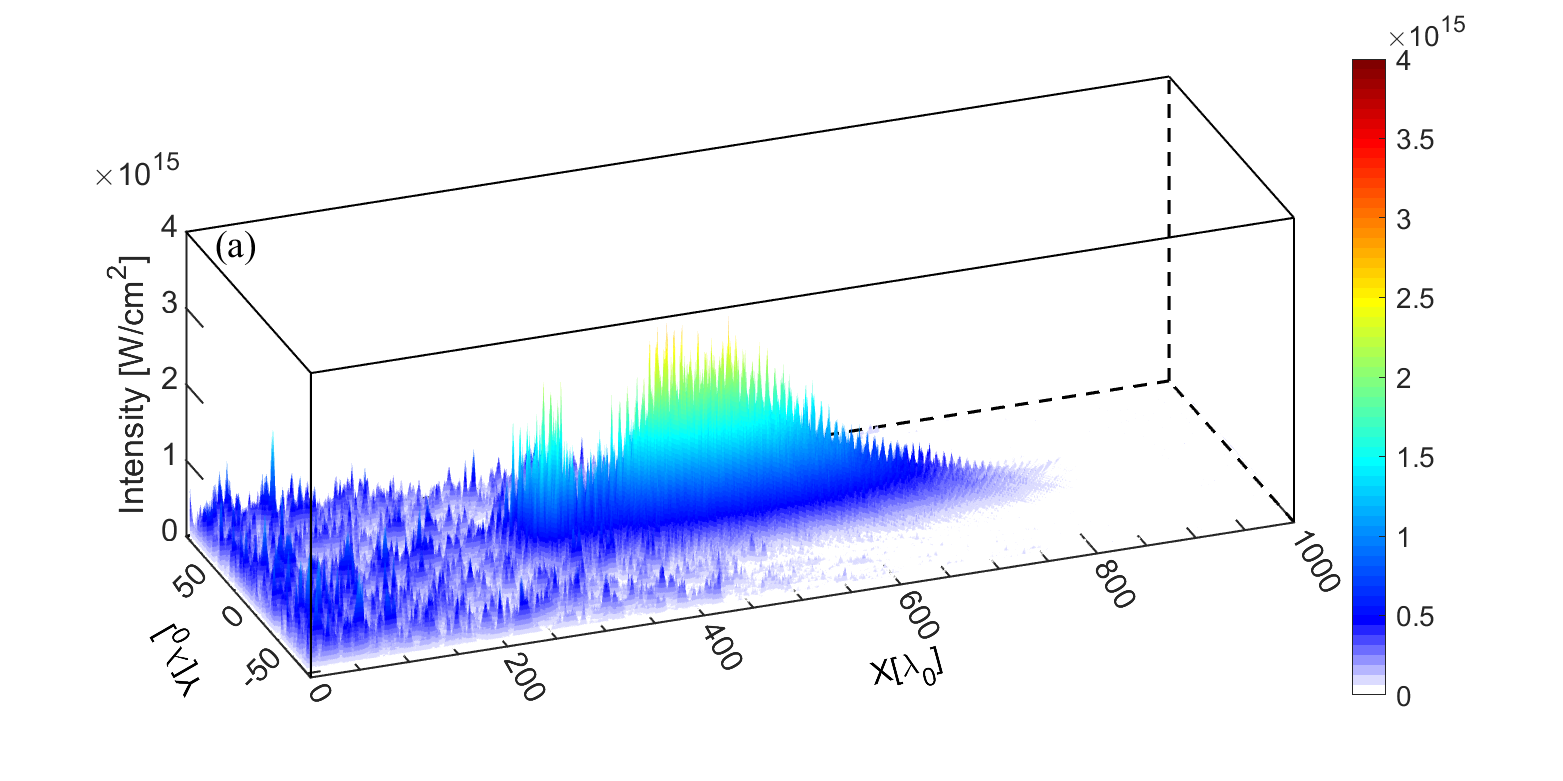}\vspace{-20pt}
    \end{minipage}
    \vspace{0.2cm}
    \begin{minipage}[t]{\linewidth}
        \centering
        \includegraphics[width=0.98\linewidth]{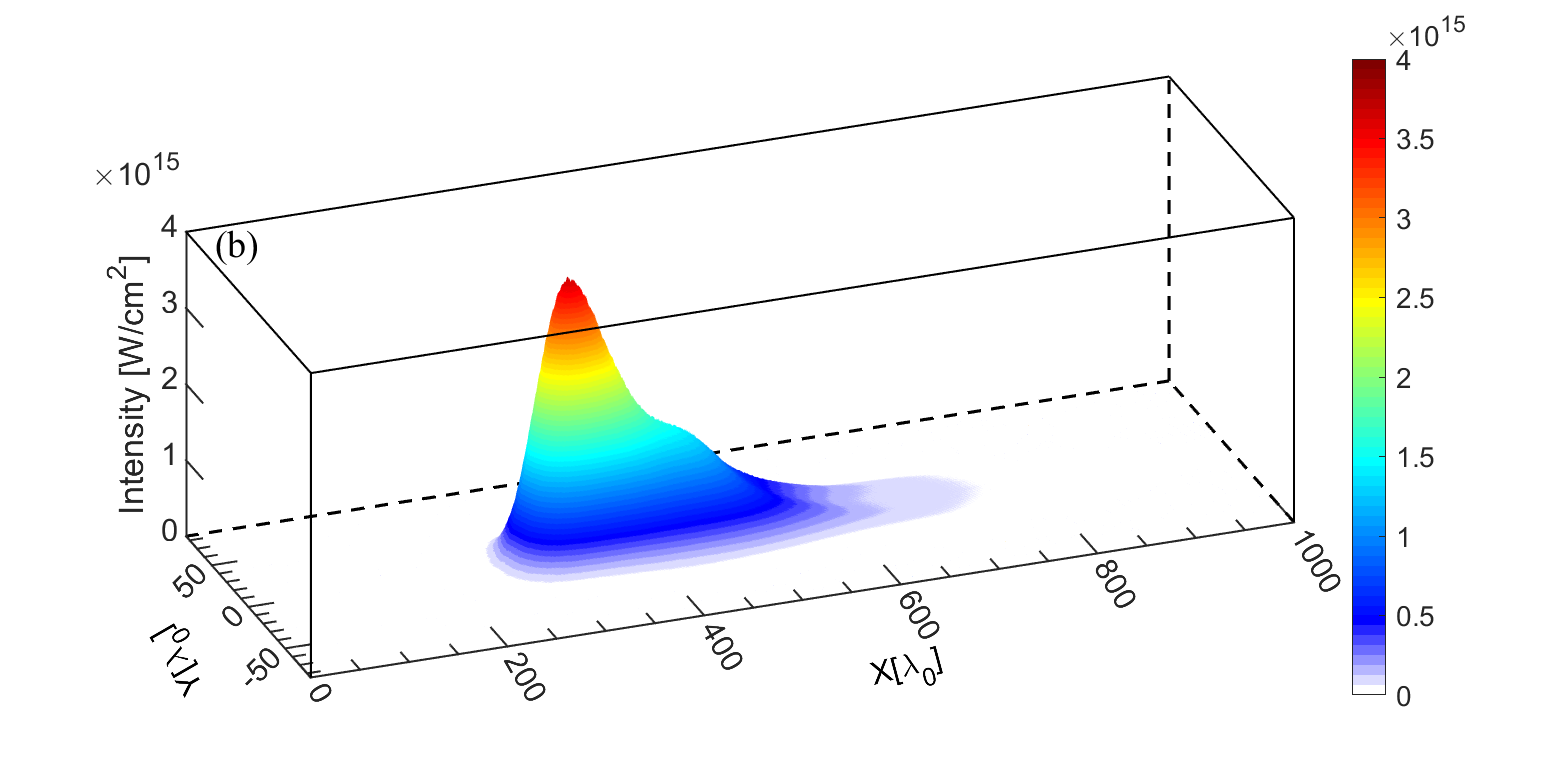}\vspace{-20pt}
    \end{minipage}
    \caption{\label{pic_2d} 2D PIC simulation results, (a) is the amplified seed laser by SC-SBS in normal electron-ion plasma,(b)is the amplified seed laser by $\mu$-wave amplification in $e^{-}$-$\mu^{-}$-$\mathrm{ion}$  plasma with $\eta = 0.2$. There are $30$ cells$/\lambda_{0}$ at x direction, and $8$ cells$/\lambda_{0}$ at y direction in PIC simulations. We also set $100$ particles per cell. }
\end{figure}

We use PIC simulations to quantitatively demonstrate the advantages of $\mu$-wave amplification over conventional amplification techniques. The amplification length is set as $1.6$ $\rm{mm}$, the plasma density is $2.34\times 10^{19}$ $\rm{cm^{-3}}$, the ions are protons, and temperatures of plasma is $T_{e}=T_{\mu}=T_{i} = 70 \rm{eV}$ ($k\lambda_{D} = 0.2$). The intensities of pump laser and seed laser are both $10^{15}$ $\rm{W/cm^2}$.

We first simulate SC-SBS amplification for $t = 1$, in Fig.~\ref{scamp} (a), the seed laser intensity reaches $8.6 \times 10^{15}$ $\rm{W/cm^2}$. One can observe that Raman-scattered light appears  in the wavenumber spectrum due to pump spontaneous instabilities. Figure~\ref{scamp}(b) shows Raman amplification,  the intensity of seed reaches to $1.1\times10^{16}$ $\rm{W/cm^2}$, but the  beam Gaussian profile distorts from pulse splitting \cite{spliting}, attributed to Langmuir wave nonlinear frequency shifts.  The distorted envelope of the Raman-amplified pulse observed in Figure~\ref{scamp}(b) is an inherent characteristic of high-intensity amplification in pure electron-ion plasmas, resulting from two primary factors: First, the high pump intensity ($10^{15}$ $\rm{W/cm^{2}}$) enables rapid growth of spontaneous Raman scattering from pump noise, whose amplitude quickly surpasses that of the seed pulse and disrupts the coherent amplification process. Second, the large-amplitude Langmuir waves driven by such intense pumps undergo significant nonlinear frequency shifts\cite{suppress5}due to particle trapping effects, leading to phase mismatch in the three-wave coupling and consequent pulse splitting.

Finally, as shown in  Fig.~\ref{scamp} (c), we implement the $\mu$-wave amplification with $\eta = 0.2$, ($i.e.$  $n_{e}:n_{\mu} = 1:4$), we observe that the intensity of seed laser increases to $1.9\times10^{16}$ $\rm{W/cm^2}$, which is almost two times of SC-SBS amplification in Fig.~\ref{scamp} (a). The reason is that the Landau damping of Langmuir wave is strong enough to suppress the  spontaneous instabilities of the pump laser. Thus, the spatial spectrum of seed laser exhibits  exceptional monochromaticity and $\mu$-wave amplification maintains the seed beam's Gaussian profile.

The density of muons and protons in $\mu$-wave amplification with $\eta =0.2$ shown in Fig.~\ref{scamp} (d) and (e), one can observe that muon density perturbations exhibit substantial growth during instability development, while the ion density remains effectively unresponsive. This phenomenon clearly indicates that ion density perturbations do not contribute to the $\mu$-wave amplification.

We also implement Raman amplification with $\eta =0.2$ (not shown here), the intensity only increases to   $4.5 \times 10^{15}$ $\rm{W/cm^2}$, the Gaussian profile of the seed beam is distorted, exhibiting similar characteristics to those shown in Fig.~\ref{scamp} (b), although the growth rate of Raman amplification is higher than that of $\mu$-wave amplification.

 The advantage of $\mu$-wave amplification is twofold: (1) The reduced effective plasma frequency leads to a lower growth rate for undesirable SRS, giving the amplification process more time to complete before instabilities develop. (2) The dramatically enhanced Landau damping of Langmuir waves actively suppresses, preventing them from consuming a significant portion of the pump energy. The spontaneous $\mu$-wave instability can indeed be excited from noise in the plasma. However, its growth from the natural thermal noise level is much slower than that of the stimulated process driven by the intense seed laser pulse.

 Although achieving an 80\% muon fraction remains experimentally challenging under current conditions, Fig.~\ref{grow_pic}(c) demonstrates that even partial electron replacement (e.g., $\eta = 0.5$ - $0.7$, corresponding to 30-50\% muon fraction) can yield meaningful enhancement of Landau damping, thereby reducing parasitic instabilities compared to standard electron-ion plasmas. While the most pronounced effects occur at $\eta =0.2$, our results reveal a continuous scaling of benefits with increasing muon fraction. This represents a crucial insight, as it indicates that future experiments could target lower, more attainable muon fractions while still observing measurable positive effects, thereby providing proof-of-principle validation for our model.

 To further demonstrate the advantages of $\mu$-wave amplification in suppressing spontaneous instabilities, we employ 2D PIC code to simulate strong-coupling Brillouin amplification in an electron-ion plasma and $\mu$-wave amplification in $e^{-}$-$\mu^{-}$-$\mathrm{ion}$ plasmas, respectively. The plasma parameters and laser intensity were kept consistent with the one-dimensional simulations in Fig.~\ref{scamp}, while the simulation box was shortened to 1000 $\lambda_{0}$.

 As shown in Fig.~\ref{pic_2d}(a) the results of SC-SBS amplification, we observe that spontaneous instabilities consume energy in the pump light entrance region, reducing the amplification efficiency. The filamentation of seed laser distorting the Gaussian profile\cite{SBS6,fila}, the growth rate of filamentation can be estimated by $\gamma_{fil} = v_{osc}^{2}\omega_{pe}^{2}/8v_{e}\omega_{pump}$. As shown in Fig.~\ref{pic_2d}(b),  $\mu$-wave amplification achieves higher seed light intensity and better preserves the Gaussian laser waveform. The growth rate of filamentation is lower in $e^{-}$-$\mu^{-}$-$\mathrm{ion}$ plasmas, because $\omega_{pe}$ is smaller when $\eta =0.2$.

%In summary, $\mu$-waves exhibit hybrid ion-acoustic/Langmuir behavior, enabling  higher amplification efficiency than SBS/Raman. This work establishes muon plasmas as a viable platform for high-fidelity laser amplification, with experimental validation becoming feasible through advancing muon sources\cite{lasermiu3}.

\section{Conclusion and discussion}\label{conclusion}

In summary, we investigate plasma waves in $e^{-}$-$\mu^{-}$-$\mathrm{ion}$ plasmas, demonstrating that $\mu$-wave exhibit ion-acoustic wave behavior at low $k\lambda_{D}$ and Langmuir wave characteristics at high $k\lambda_{D}$. Through theoretical analysis and PIC simulations, we quantify the instability growth rate in these systems. $\mu$-wave amplification maintains excellent seed beam Gaussian profile fidelity and spectral purity, even at high pump intensities.

This work establishes a new framework for plasma physics involving multi-species negatively charged particles.   The model of instabilities  providing essential theoretical tools for extreme-condition plasma manipulation. The practical realization of the $\mu$-wave amplification scheme proposed here ultimately depends on the continued development of muon source technology. The key challenge lies in compressing the muon phase-space density via techniques such as ionization cooling to achieve a particle density high enough to form a plasma. As muon generation and plasma diagnostics advance, experimental verification will bridge high-energy and plasma physics, opening new research directions.

\section*{Acknowledgements}

 This work was supported by the Scientific Research Foundation for High-level Talents of Anhui University of Science and Technology (Grant No. 2022yjrc106), National Natural Science Foundation of China (Grant Nos. 12405277 and 12475237), Anhui Provincial Natural Science Foundation (Grant No. 2308085QA25), the Science Challenge Project (No.TZ2025012) and the Fund of National Key Laboratory of Plasma Physics (Grant No. 6142A04230103).

%\bibliographystyle{apsrev4-1}
%\bibliography{ClassicSpinModels}

\begin{thebibliography}{100}
\newcommand{\DOI}[1]{doi: \href{https://doi.org/#1}{#1}}


\bibitem{muonic1} P. A. Souder,  et al.,  Phys. Rev. Lett. {\bf 34}, 1417 (1975). (\DOI{10.1103/PhysRevLett.34.1417})
\bibitem{muonic2} P. A. Souder,  et al.,  Phys. Rev. A. {\bf 22}, 33 (1980). (\DOI{10.1103/PhysRevA.22.33})
\bibitem{muonic3} Y. G. Jiang, X. N. Wang, X. F. Lan, Y. S. Huang, Phys. Plasmas {\bf 29}, 103110 (2022).
(\DOI{10.1063/5.0107458})


\bibitem{muonic4} Pan-Fei Geng, Min Chen, Zheng-Ming Sheng, Phys. Plasmas {\bf 31}, 023109 (2024).
(\DOI{10.1063/5.0189289})


\bibitem{muonic5} K. N. Borozdin, G. E. Hogan, C. Morris, W. C. Priedhorsky, A. Saunders, L. J. Schultz, and M. E. Teasdale, Nature {\bf 422}, 277 (2003).(\DOI{10.1038/422277a})


\bibitem{muonic6} S. J. Blundell,  Contemp. Phys. {\bf 40}, 175 (1999).(\DOI{10.1080/001075199181521})
\bibitem{muonic7} J. Blas, J. Gu, and Z. Liu, Phys. Rev. D {\bf 106}, 073007 (2022).(\DOI{10.1103/PhysRevD.106.073007})

\bibitem{muonic8} S. Aritome, et al. Phys. Rev. Lett. {\bf 134}, 245001 (2025).(\DOI{10.1103/PhysRevLett.134.245001})


\bibitem{cosmic1}  Morishima K.,  et al.  Nature {\bf 552}, 387-401 (2017).(\DOI{10.1038/nature24647})

\bibitem{cosmic2} T. Pierog and K. Werner, Phys. Rev. Lett. {\bf 101}, 171101 (2008).(\DOI{10.1103/PhysRevLett.101.171101})

\bibitem{cosmic3} A. Aab et.al., Phys. Rev. Lett. {\bf 126}, 152002 (2021).(\DOI{10.1103/PhysRevLett.126.152002})

 \bibitem{labmiu1} A. Czrne, S.F.J. Cox, G. H. Eaton and C.A. Scott, Hyperfine Interact. {\bf 65}, 1175-1181 (1991).(\DOI{10.1007/BF02397776})
 \bibitem{labmiu2}  K. Shimomura, et al. Interactions, {\bf 245}, 31 (2024).(\DOI{10.1007/s10751-024-01863-8})

 \bibitem{labmiu3}  Wang, Z. et al.  Phys. Rev. Accel. Beams {\bf 27}, 010101 (2024).(\DOI{10.1103/PhysRevAccelBeams.27.010101})

 \bibitem{lasermiu1} L. Calvin, P. Tomassini, D. Doria,D. Martello, R. M. Deas and G. Sarri, Front. Phys. {\bf 11}, 1177486 (2023). (\DOI{10.3389/fphy.2023.1177486})

 \bibitem{lasermiu2}  A. I. Titov, B. K$\ddot{a}$mpfer, and H. Takabe, Phys. Rev. ST Accel. Beams {\bf 12}, 111301 (2009). (\DOI{10.1103/PhysRevSTAB.12.111301})

 \bibitem{lasermiu4} T. Ziegler,I. G$\ddot{o}$the,  S. Assenbaum et.al., Nat. Phys. {\bf 20} 1211-1216 (2023).(\DOI{10.1038/s41567-024-02505-0})

 \bibitem{lasermiu3}  F. Zhang, L. Deng, Y. Ge, et al., Nat. Phys. Advance online publication. (2025). (\DOI{10.1038/s41567-025-02872-2})

 \bibitem{cooling}  M. Bogomilov et al., Nature, {\bf 578},53-59 (2020).(\DOI{10.1038/s41586-020-1958-9})


\bibitem{e_p_2016}M. R. Edwards, N. J. Fisch, and J. M. Mikhailova, Phys. Rev. Lett. {\bf 116}, 015004 (2016).(\DOI{10.1103/PhysRevLett.116.015004})

\bibitem{e_p_2017}  F. Schluck, G. Lehmann, and K. H. Spatschek, Phys. Rev. E, {\bf 96},053204 (2017).(\DOI{10.1103/PhysRevE.96.053204})

\bibitem{chen1} Y. Chen, C. Y. Zheng, Z. J. Liu, L. H. Cao, Q. S. Feng, andC.Z.Xiao, Plasma Phys. Controlled Fusion {\bf 62}, 105020 (2020). (\DOI{10.1088/1361-6587/ab98df})

 \bibitem{malkin1}V. M. Malkin, G. Shvets and  N. J. Fisch,  Phys. Rev. Lett. {\bf 82}, 4448 (1999). (\DOI{10.1103/PhysRevLett.82.4448})
 \bibitem{SRS1} R. M. G. M. Trines, F. Fi$\acute{u}$za, R. Bingham, R. A. Fonseca, L. O. Silva, R. A. Cairns and P. A. Norreys, Nature Phys. {\bf 7}, 87 (2011).(\DOI{10.1038/nphys1793})
 \bibitem{SRS2} J. Ren, W. Cheng, S. Li and S. Suckewer, Nature Phys. {\bf 3}, 732 (2007).(\DOI{10.1038/nphys717})
 \bibitem{SRS3} Z. Toroker, V.M. Malkin and  N. J. Fisch, Phys. Rev. Lett. {\bf 109}, 085003 (2012).(\DOI{10.1103/PhysRevLett.109.085003})



\bibitem{Riconda2d} C. Riconda, S. Weber, L. Lancia, J.-R. Marqu$\grave{e}$s, G. Mourou and J. Fuchs, Phys. Plasmas Control. Fusion {\bf 57},  014002(2015) (\DOI{10.1088/0741-3335/57/1/014002})


\bibitem{SBS2} M. R. Edwards, Q. Jia, J. M. Mikhailova and N. J. Fisch, Phys. Plasmas {\bf 23}, 083122 (2016).(\DOI{10.1063/1.4961429})


\bibitem{SBS4} A. A. Andreev, C. Riconda, V. T. Tikhonchuk and  S. Weber, Phys. Plasmas {\bf 13}, 053110 (2006).(\DOI{10.1063/1.2201896})
\bibitem{SBS5} L. Lancia, J.-R. Marqu$\grave{e}$s, M. Nakatsutsumi, C. Riconda, S. Weber, S. H$\ddot{u}$ller, A. Man$\check{c}$i$\acute{c}$,1 P. Antici, V. T. Tikhonchuk, A. H$\acute{e}$ron, P. Audebert and J. Fuchs, Phys. Rev. Lett. {\bf 104}, 025001 (2010).(\DOI{10.1103/PhysRevLett.104.025001})

 \bibitem{SBS7}  Z. M. Zhang, B. Zhang, W. Hong, Z. G. Deng, J. Teng, S. K. He, W. M. Zhou, and Y. Q. Gu, Phys. Plasmas, {\bf 24}, 113104 (2017).(\DOI{10.1063/1.4999651})
 \bibitem{SBS10} M. Chiaramello, F. Amiranoff, C. Riconda and S. Weber, Phys. Rev. Lett. {\bf 117}, 235003 (2016).(\DOI{10.1103/PhysRevLett.117.235003})
 \bibitem{Amir} F. Amiranoff, C. Riconda, M. Chiaramello, L. Lancia, J. R. Marqu$\grave{e}$s and  S. Weber, Phys. Plasmas {\bf 25}, 013114 (2018).(\DOI{10.1063/1.5019374})


  \bibitem{suppress1} V. M. Malkin and  N. J. Fisch, Phys. Plasmas, {\bf 24}, 4698 (2001)(\DOI{10.1063/1.1400791})

  \bibitem{suppress2} D. S. Clarka and N. J. Fisch, Phys. Plasmas, {\bf 9}, 2772 (2002)(\DOI{10.1063/1.1471515})
  \bibitem{suppress3} D. S. Clarka and N. J. Fisch, Phys. Plasmas, {\bf 10}, 4837 (2003)(\DOI{10.1063/1.1625939})

  \bibitem{suppress4} D. S. Clark and N. J. Fisch, Phys. Plasmas {\bf 10}, 3363 (2003).(\DOI{10.1063/1.1590667})



  \bibitem{gas} Y. Ping,W. F, Cheng, and S. Suckewer, Phys. Rev. Lett., {\bf 92}, 175007 (2004).(\DOI{10.1103/PhysRevLett.92.175007})
   \bibitem{fly} D. Turnbull, S. Bucht, A. Davies, D. Haberberger, T. Kessler, J. L. Shaw, and D. H. Froula, Phys. Rev. Lett., {\bf 120}, 024801 (2018).(\DOI{10.1103/PhysRevLett.120.024801})

   \bibitem{spliting}  Q. Chen, Z. Wu, L. Johnson, D. Gordon, P. Sprangle, and S. Suckewer, Phys. Plasmas, {\bf 24}, 123113 (2017).(\DOI{10.1063/1.5009553})

   \bibitem{suppress5}   V. M. Malkin, G. Shvets, and N. J. Fisch, Phys. Rev. Lett. {\bf 84}, 1208 (2000).
(\DOI{10.1103/PhysRevLett.84.1208})

   \bibitem{suppress6} G. Lehmann and K. H. Spatschek, Phys. Plasmas, {\bf 22}, 043105 (2015). (\DOI{10.1063/1.4916958})

   \bibitem{suppress7} F. Schluck, G. Lehmann, and K. H. Spatschek, Phys. Plasmas,  {\bf 22}, 093104 (2015)(\DOI{10.1063/1.4929859})
   \bibitem{nosie3} Q. Jia, I. Barth, M. R. Edwards, J. M. Mikhailova,and N. J. Fisch, Phys. Plasmas {\bf 23}, 053118 (2016).(\DOI{10.1063/1.4951027})


   \bibitem{SBS6} Y. Chen, C. Y. Zheng, Z. J. Liu, L. H. Cao, and C. Z. Xiao, Phys. Rev. E {\bf 107}, 015204 (2023).(\DOI{10.1103/PhysRevE.107.015204})

   \bibitem{nosie1} S. Weber, C. Riconda, L. Lancia, J.-R. Marqu$\grave{e}$s, G. A. Mourou and J. Fuchs, Phys. Rev. Lett. {\bf 111}, 055004 (2013).(\DOI{10.1103/PhysRevLett.111.055004})
   \bibitem{nosie2}  C. Riconda, S. Weber, L. Lancia, J. R. Marques, G. A. Mourou, and J. Fuchs, Phys. Plasmas {\bf 20}, 083115 (2013).(\DOI{10.1063/1.4818893})

   \bibitem{twoele} W. D. Jones, A. Lee, S. M. Gleman, and H. J. Doucet, Phys. Rev. Lett. {\bf 20}, 1349 (1975).(\DOI{10.1103/PhysRevLett.35.1349})

   \bibitem{twoele2} S. P. Gary and R. L. Tokar, Phys. Fluids {\bf 28}, 2439 (1985).(\DOI{10.1063/1.865250})

   \bibitem{twoele3} R. L. Mace, G. Amery, and M. A. Hellberg, Phys. Plasmas {\bf 6}, 44 (1999).  (\DOI{10.1063/1.873256})



    \bibitem{xiehs}  H. S. Xie, Phys. Plasmas {\bf 20}, 092125 (2013).(\DOI{10.1063/1.4822332})


    \bibitem{fly2}  Z. Wu, Y. Zuo, Z. Zhang, et.al. Phys. Rev. E, {\bf 106}, 035209 (2022).(\DOI{10.1103/PhysRevE.106.035209})


     \bibitem{fila}  Z. Li, Y. L. Zuo, J. Q. Su, and S. H. Yang, Phys. Plasmas, {\bf 26}, 093102 (2019).
(\DOI{10.1063/1.5094513})
\end{thebibliography}
%\noindent\makebox[\columnwidth]{\rule{\columnwidth}{0.5pt}}

\end{document}